# Quantum oscillations in the parent magnetic phase of an iron arsenide high temperature superconductor

**Suchitra E. Sebastian[1*], J. Gillett[1], N. Harrison[2], P. H. C. Lau[1], D. J. Singh[3], C. H. Mielke[2] & G. G. Lonzarich[1]**

1. Cavendish Laboratory, Cambridge University, JJ Thomson Avenue, Cambridge CB3 OHE, U.K.

2. NHMFL, Los Alamos National Laboratory, MS E536, Los Alamos, NM 87545, U.S.A.

3. Materials Science and Technology Division, Oak Ridge National Laboratory, Oak Ridge, TN 37831, U.S.A.

We report quantum oscillation measurements in $SrFe_2As_2$ - which is an antiferromagnetic parent of the iron-arsenide family of superconductors - known to become superconducting under doping and the application of pressure. The magnetic field and temperature dependences of the oscillations between 20 and 55 T in the liquid helium temperature range suggest that the electronic excitations are those of a Fermi liquid. We show that the observed Fermi surface comprising small pockets is consistent with the formation of a spin-density wave. Our measurements thus demonstrate that high $T_c$ superconductivity can occur on doping or pressurizing a conventional metallic spin-density wave state.

[*] Correspondence to be addressed to: S.E.S. (suchitra@phy.cam.ac.uk)

## 1. Introduction

Knowing the nature of elementary excitations and their interplay with the Fermi surface is essential for understanding the origin of unconventional pairing in high temperature superconductors [ref. 1,2]. These are rendered inaccessible in the parent antiferromagnetic phases of layered cuprate superconductors, since their strongly coupled nature precludes the existence of a Fermi surface [ref. 3]. In contrast, the electrically conducting nature of the parent phase in the recently discovered FeAs-layered family of high temperature superconductors [ref. 4], leaves open the possibility of fermionic excitations [res. 5-10]. In this paper we report the results of quantum oscillation experiments in very strong magnetic fields on the parent magnetic compound $SrFe_2As_2$ of a high temperature superconductor, revealing a Fermi surface comprising tiny pockets of carriers. Our observation of quantum oscillations in the parent FeAs family of high temperature superconductors is a breakthrough in showing that high temperature superconductivity can emerge on doping a conventional metallic antiferromagnetic state. The small sizes of the Fermi surface pockets, corresponding to ~ 1 % of the original Brillouin zone area, are suggestive of an itinerant inter-band antiferromagnetic order parameter that destroys almost the entire paramagnetic Fermi surface.

## 2. Experimental details and results

$A Fe_2As_2$ ($A$=Sr, Ba) compounds belong to the recently discovered FeAs family of high $T_c$ superconductors [refs. 4, 11, 12]. These materials form in the body-centered tetragonal $ThCr_2Si_2$ structure [ref. 13], now known to be common to several families of unconventional superconducting compounds [refs. 14, 15]. At $T_N \approx 200$ K, $SrFe_2As_2$ exhibits a phase transition [ref. 13] thought to be associated with antiferromagnetism within the FeAs layers, accompanied by a structural transition which lowers the crystal symmetry [refs. 16, 17]. On suppressing magnetic order by pressure [ref. 18], or by chemical doping in a similar fashion to that in undoped CuO-layers in cuprate high $T_c$ superconductors, superconductivity is found to emerge. An optimal transition temperature of $T_c \approx 37$ K is reached on substituting 40 % of the Sr atomic sites with K [ref. 19], corresponding to a doped hole concentration of $p \approx 0.20$ per Fe— similar to that in hole-doped cuprates.

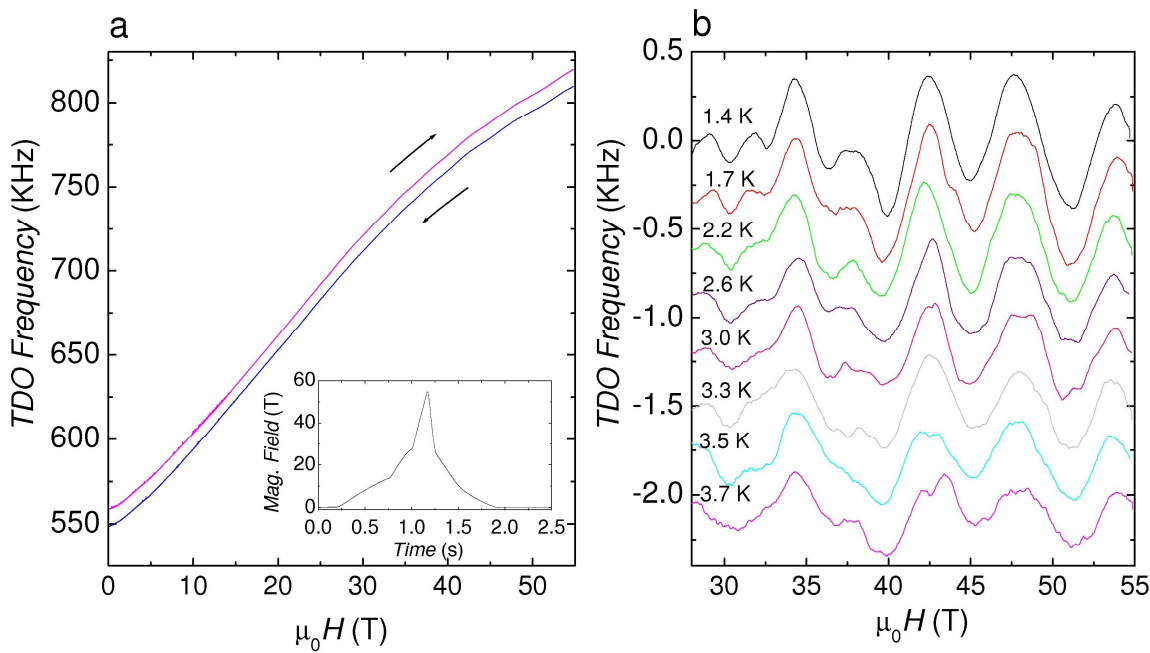

**Fig. 1. SrFe$_2$As$_2$ experimental data. a,** The tunnel diode oscillator (TDO) resonance frequency (lowered by mixing) measured on the rising and falling magnetic field is shown (offset for clarity) – a faint outline of oscillations is perceptible. The inset shows the $\mu_0H$ versus time profile of the applied magnetic field. **b,** Quantum oscillations in the TDO frequency obtained on performing a background polynomial subtraction from (a) are shown at a series of different temperatures (also offset for clarity).

The ratio of resistivity between room and liquid $^4$He temperatures ($\sim 8$) in our single crystalline samples of SrFe$_2$As$_2$ is relatively low compared to good metals, suggesting that quantum oscillations could be challenging to observe. We therefore utilize strong magnetic fields provided by the 60 tesla motor-generator-driven magnet at Los Alamos and the 45 tesla continuous field magnet at Tallahassee combined with the radio frequency contactless conductivity technique, which has recently proven to be a powerful tool for Fermi surface studies in high temperature superconducting cuprates [refs. 20, 21]. Figure 1a shows the results of such a quantum oscillation experiment performed on $SrFe_2As_2$ in the motor generator magnet, supplemented by quantum oscillation experiments in the hybrid magnet (inset). Quantum oscillations are observed through a magnetic field-dependent change in the resonance frequency of a tunnel diode oscillator (TDO) circuit to which a single crystalline sample of $SrFe_2As_2$ is coupled inductively (see Appendix A). On subtracting a fitted polynomial background in Fig. 1b, Fourier transformation of the quantum oscillation waveform in inverse magnetic field in Fig. 2a reveals three prominent frequencies $F_\alpha = 370 \pm 20$ T, $F_\beta = 140 \pm 20$ T, and $F_\gamma = 70 \pm 20$ T. Applying the Onsager expression $F = (\hbar/2\pi e)A_k$ [ref. 22], these are found to correspond to pockets ($\alpha$, $\beta$ and $\gamma$) with $k$-space areas occupying 1.38, 0.52, and 0.26 % of the paramagnetic basal plane Brillouin zone respectively. Measurements made of the same crystal on rotating the magnetic field by an angle $\theta$ within the **ac** plane in the inset of Fig. 2b indicate that the Fermi surface topology corresponding to each of the pockets is approximately consistent with a prolate ellipsoid of revolution with its principal axes along the **c** axis. On extracting the temperature dependence of the oscillation amplitude from the Fourier transforms and fitting to the Lifshitz-Kosevich thermal broadening term [ref. 22] (shown in Fig. 2b), the effective masses of the $\alpha$ and $\beta$ pockets are found to be $m^*_\alpha = 2.0 \pm 0.1 \ m_e$ and $m^*_\beta = 1.5 \pm 0.1 \ m_e$ (where $m_e$ is the free electron mass).

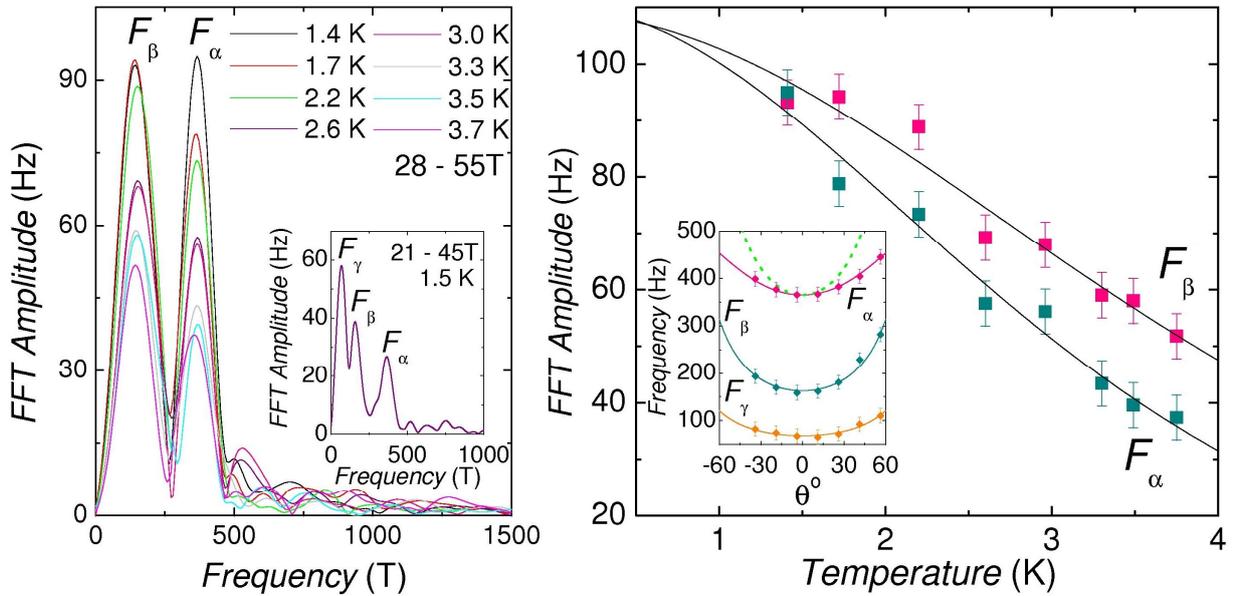

**Fig. 2. Analysis of the SrFe₂As₂ quantum ocillations. a,** Fourier transform of the quantum oscillations performed over the interval $28 < \mu_0 H < 55$ T at many different temperatures in the motor-generator magnet and over the interval $21 < \mu_0 H < 45$ T in the hybrid magnet (inset), yielding three prominent frequencies that we denote $F_\alpha$, $F_\beta$ and $F_\gamma$. **b,** A fit of the experimental temperature-dependent amplitudes to the Lifshitz-Kosevich thermal damping term $A_T = A_0 \, X/\sinh X$, where $X = 14.69 \, m^*T/B$ ($m^*$ relative to the free electron mass $m_e$). Adherence to the functional form of $A_T$ is consistent with the existence of elementary excitations governed by a conventional Fermi-Dirac distribution function[11]. Additional measurements of the frequencies for **H** rotated away from **c** by an angle θ within the **ac** plane are shown in the inset. The solid lines are fits of $F$ versus θ to ellipsoids of revolution $F = F_0 \left( \cos^2\theta + \dfrac{1}{r^2}\sin^2\theta \right)^{-\frac{1}{2}}$ (yielding $r_\alpha \approx 1.4$, $r_\beta \approx 6.1$, and $r_\gamma \approx 3.3$), while the dotted line is the expected θ-dependence for the innermost warped hole cylinder of the unreconstructed Fermi surface shown in Fig. 3a.

## 3. Discussion

The generic features of the Fermi surface of this family of materials are similar to those reported in the oxy-pnictide compounds [refs. 23, 24]. The calculated Fermi surface of paramagnetic SrFe₂As₂ (see Fig. 3a), which may also been observed by angle-resolved photoemission spectroscopy (ARPES) [ref. 25], consists of three concentric warped cylindrical hole pockets centred at Γ (0,0,0) and two concentric warped cylindrical electron pockets situated at M (π,π,0) (see appendix B). On comparing our observed pockets with those of paramagnetic SrFe₂As₂ in Fig. 3a, the smaller observed frequencies $F_\beta$ and $F_\gamma$ do not correspond to any of the predicted features in the unreconstructed Fermi surface. While the larger observed cross-section corresponding to $F_\alpha$ is comparable in size to the narrowest cross-section of the innermost hole-like cylinder situated at $\Gamma$, the expected $\theta$-dependence of this section (Fig. 2a inset) does not follow that observed for $F_\alpha$. Furthermore, the orbitally averaged Fermi velocity $v_F = \sqrt{2\hbar eF}\big/m^* \approx 6 \times 10^4$ ms$^{-1}$ for both $\alpha$ and $\beta$ pockets is reduced compared to that $v_h \approx 1.7 \times 10^5$ ms$^{-1}$ and $v_e \approx 1.1 \times 10^5$ ms$^{-1}$ averaged over the calculated hole and electron sections shown in Fig. 3a. Such discrepancies suggest that the system likely exhibits a form of Fermi surface reconstruction resulting from staggered spin or charge ordering at low temperatures.

The details of the Fermi surface reconstruction depend on how neighbouring Fe spins are aligned within the unit cell. Spin-density wave order is suggested by elastic neutron scattering experiments that have recently identified a wavevector $\mathbf{Q} = (\pi,\pi,0)$ for antiferromagnetic ordering in both LaO$_{1-x}$F$_x$FeAs [ref. 5] and SrFe$_2$As$_2$ [ref. 26]. In such a case, the antiferromagnetism is of an itinerant inter-band nature qualitatively similar to that in elemental Cr [ref. 27]. The reconstructed Fermi surface corresponding to the reduced magnetic Brillouin zone is thus determined by the 'nesting' of similarly sized electron and hole pockets as shown schematically in Fig. 3a. A consequence of such ordering for FeAs layers is that the fourfold symmetry of the Fermi surface is lost [ref. 5], with the Fe moments being staggered parallel to $\mathbf{Q}$ and collinear orthogonal to $\mathbf{Q}$ as shown in Fig. 3.

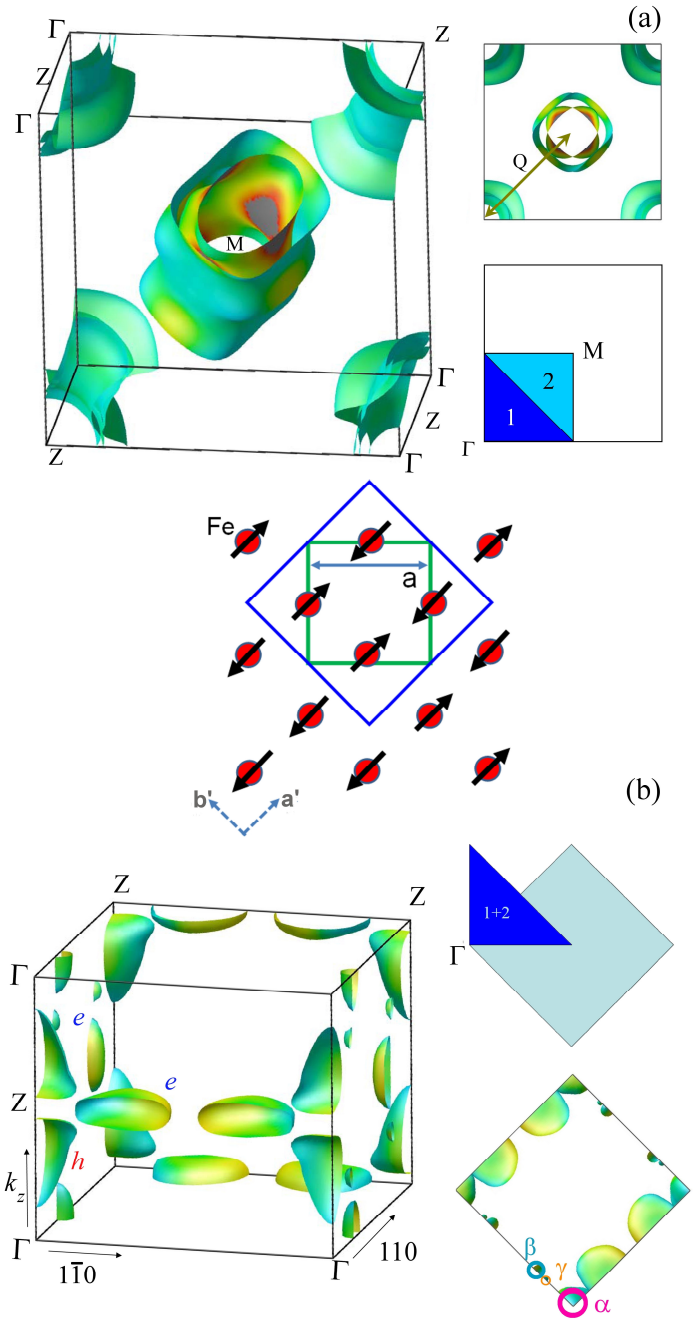

**Fig. 3. The Fermi surface of SrFe₂As₂. a,** The paramagnetic (unreconstructed) Fermi surface of SrFe₂As₂ calculated using the local density approximation (LDA) method (see appendix B), together with a top-down view showing the relative sizes of the sections within the Brillouin zone. For an antiferromagnetic wavevector of $\mathbf{Q} = (\pi, \pi, 0)$, the two outer most hole cylinders at $\Gamma$ 'nest' with the two electron cylinders at $\mathbf{M}$, resulting in a significant reconstruction of the Fermi surface upon folding of the regions labelled 1 and 2. **b,** The reconstructed Fermi surface resulting from folding together with a top-down view of the antiferromagnetic Brillouin zone. The sizes of the experimentally observed $\alpha$, $\beta$ and $\gamma$ pockets are shown for comparison. The colour graduation of the Fermi surface represents the local Fermi velocity with red and blue corresponding to high and low velocities respectively. The schematic between (a) and (b) shows the ordering of Fe moments staggered parallel to $\mathbf{Q}$ and collinear orthogonal to $\mathbf{Q}$, with the antiferromagnetic unit cell depicted in blue. The crystalline axis of the tetragonal unit cell is denoted as $a$, while the axes of the orthorhombic unit cell are denoted by $a'$, $b'$.

Figure 3b shows an example of how the Fermi surface in Fig. 3a is reconstructed by **Q**, yielding small pockets of varying size. While the details of the Fermi surface depend on the magnitude of the staggered Fe moments and the precise positions of the As atoms within the unit cell (see appendix B), qualitative agreement with the observed quantum oscillation frequencies can be inferred. The tear-shaped hole pockets located along $\Gamma Z$ are what remains of the innermost hole cylinder in Fig. 3a, and have a comparable Fermi surface cross-section and volume to our observed $\alpha$ pocket. The remainder of the calculated Fermi surface consists of ellipsoidal electron pockets whose total volume equals that of the hole pockets so as to preserve overall charge neutrality within the reduced antiferromagnetic Brilouin zone. The two smallest electron pockets are of comparable dimensions to the observed $\beta$ and $\gamma$ pockets, while the largest is as yet undetected in quantum oscillation experiments.

The contribution of the experimentally observed Fermi surface to the electronic heat capacity $C = \gamma T$ is estimated by comparison with the qualitatively similar reconstructed Fermi surface sections obtained from bandstructure calculations (Fig. 3b). The electron and hole pockets shown in Fig. 3b have orbitally averaged Fermi velocities higher by a factor of $\approx 1.85$ compared to those measured, suggesting a renormalization of the quasiparticle effective mass. Their renormalized contribution to the electronic heat capacity, therefore, yields $1.85 \times \gamma_{band} = \gamma_{renorm} = 3.5 \text{ mJ}^{-1}\text{mol}^{-1}\text{K}^{-2}$. This value of $\gamma_{renorm}$ is very similar to that $\gamma_C \approx 3.3 \text{ mJmol}^{-1}\text{K}^{-2}$ extracted from heat capacity experiments [ref. 28]. This level of agreement suggests that we can rule out the existence of the large sections of Fermi surface at low temperatures, as predicted for the paramagnetic Brillouin zone in Fig. 3a.

## 4. Significance

When considered in the framework of an inter-band spin-density wave, the small pocket sizes and low experimentally determined Fermi velocities indicate the ordering in $SrFe_2As_2$ to be more strongly coupled in nature than in elemental Cr [ref. 27]. A mere incremental enhancement of the coupling (by $\Delta\varepsilon = \hbar e F / m^* \sim 13$ meV) would cause $SrFe_2As_2$ to become completely gapped and insulating like the antiferromagnetic undoped cuprate superconductors. The metallic nature of the FeAs parent family of high temperature superconductors – enabling us to observe quantum oscillations – is therefore rather fortuitous. Given that the pockets occupy a tiny fraction of the Brillouin zone, a small (~ 1 %) doping of holes onto the FeAs layers from K substitution for Sr would cause the electron pockets to become completely depopulated, resulting in a Fermi surface consisting entirely of holes for a sufficiently large antiferromagnetic gap. A strong suppression of $T_N$ by doping or by pressure, however, must eventually cause the electron sections to reappear preceding a quantum critical point. The sensitivity of the electron and hole Fermi surfaces of $Sr_{1-x}K_xFe_2As_2$ with doping or pressure could provide a unique opportunity to pinpoint the location in $k$-space where Cooper pairs begin to form [ref. 10].

While the ultimate relevance of these results on the pnictide family of superconductors to high $T_c$ cuprates will depend on the nature of the pairing mechanism and Cooper pair wavefunction symmetry in the two systems [ref. 9], a comparative study of the elementary excitations is informative. Mott physics, implying strong correlations and non-fermionic excitations, remains a central point of discussion for unconventional pairing in high $T_c$ [refs. 1,2,9,29]. Our present results indicate that strong correlations in the parent high $T_c$ antiferromagnetic phase $SrFe_2As_2$ are no obstacle to the establishment of conventional metallic behaviour. Importantly, Mott physics is not a requisite characteristic of the parent phase for high $T_c$ superconductivity to develop.

Note added in proof: Recent ARPES measurements in another parent material $BaFe_2As_2$ in the pnictide family of superconductors [32] report an electronic structure comprising large Fermi surface sections, in contrast to the small reconstructed Fermi surface indicated by quantum oscillation measurements reported here.

## Acknowledgements

This work was supported by the UK EPSRC, the National Science Foundation, and the Department of Energy (US). S.E.S acknowledges support from the Institute for Complex Adaptive Matter and Trinity College (Cambridge University). We acknowledge assistance from J. R. Cooper, S. K. Goh, E. C. T. O'Farrell, M. Sutherland, C. Zentile, P. Alireza, P. B. Littlewood, Z. Feng, M. Grosche, S. C. Riggs, C. D. Batista, V. Zapf, A. Paris, D. Roybal, M. Gordon, and D. Rickel. S.E.S thanks I. R. Fisher, P. C. Canfield, and Z. Fisk for the tradition of crystal growth.

## Appendix A: Experimental methods

Plate-like single crystals of $SrFe_2As_2$ are grown by conventional flux-growth techniques [ref. 30] using Sn flux and starting elements of Sr, Fe and As with purities greater than 99.99 %. A pristine sample of dimensions $\sim 1 \times 2 \times 0.1$ mm$^3$ (preselected to have minimal flux inclusion) is mounted with its tetragonal $\mathbf{c}$ axis aligned parallel to the magnetic induction $\mathbf{B} \approx \mu_0\mathbf{H}$. Its planar face is attached to a 5 turn compensated coil of diameter $\sim 0.7$ mm that forms part of a tunnel diode oscillator (TDO) circuit. The oscillator resonates at a frequency of $\sim 46$ MHz in the absence of an applied magnetic field, dropping by $\sim 300$ kHz in 55 T in response to the magnetoresistivity of the sample. A change in the magnetoresistivity of the sample changes its skin depth, which in turn changes the coil inductance and resonance frequency. The quantum oscillations therefore originate from the Shubnikov-de Haas (SdH) effect. We see no evidence for an SdH effect from Sn inclusions, for which comparable frequencies would have a corresponding effective masses of ~ 0.1 $m_e$ [ref. 31].

For data acquisition purposes, the ~ 46 MHz TDO frequency is mixed down to < 1 MHz in two stages whereupon it is digitized at a rate of 20 MHz during the 2.5 s up and down sweep of the magnetic field. The sample is immersed in liquid $^4$He throughout the experiments with the temperature controlled through the vapour pressure and measured using a calibrated Cernox thermometer.

**Appendix B: Bandstructure calculations**

Local density approximation (LDA) bandstructure calculations are performed for the antiferromagnetic (AF) and paramagnetic Brillouin zones. While the structural and magnetic transitions are coincident in SrFe$_2$As$_2$, in related FeAs based compounds they are found to be separated in temperature [ref. 13]. For simplicity we assume the tetragonal unit cell throughout the calculations. Figure B1 shows the bandstructure for the paramagnetic phase. The latter Brillouin zone is halved in size as a consequence of nesting (see Fig. 3a). Ordering, in which the AF Fe sheets are stacked antiferromagnetically along the **c**-axis, yields a ground state energy 10 meV/Fe lower than that when they are stacked ferromagnetically. Similar calculations for BaFe$_2$As$_2$ yield a lower 3 meV/Fe energy for the AF **c**-axis stacking as well, which may be consistent with the lower $T_N$ in that compound. The primitive unit cell, corresponding to this AF ordering, contains 4 atoms, with orthorhombic symmetry. The AF Brillouin zone is determined by folding M onto Γ, as shown in Fig. 3. The calculated AF Fermi surface is largely gapped away, with only small sections remaining.

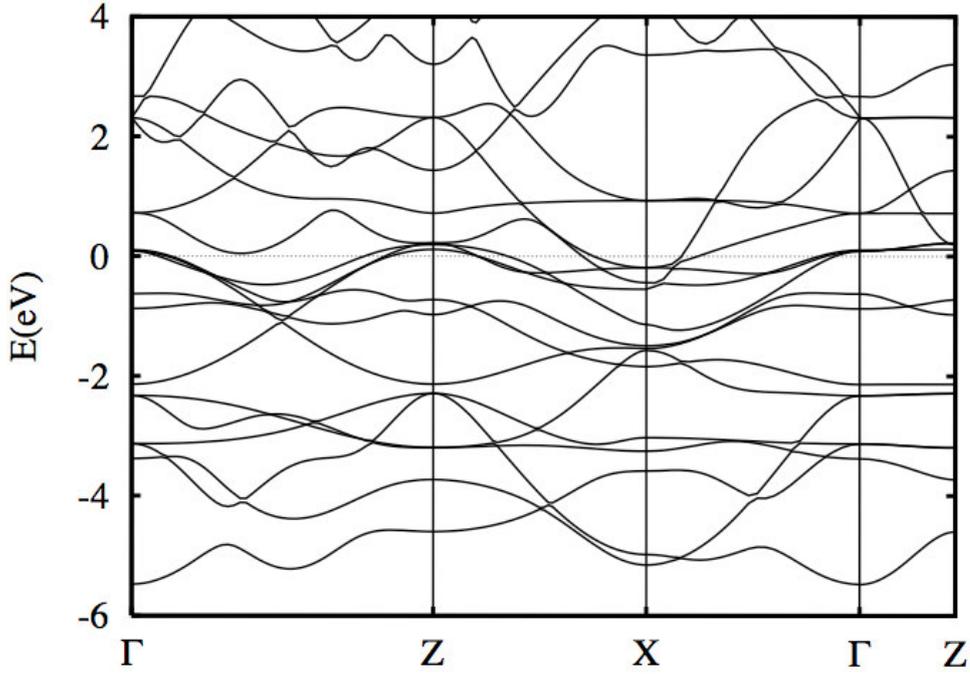

**Fig. B1.** LDA band structure of paramagnetic SrFe$_2$As$_2$ with the experimental crystal structure. The bands are shown in directions in the basal plane and along $k_z$. The long $\Gamma$Z direction is from (0,0,0) to (2$\pi$,0,$\pi$) which is perpendicular to kz in the body-centered tetragonal unit cell. The X point is ($\pi$,$\pi$,$\pi$), and the short $\Gamma$Z direction is from (0,0,0) to (0,0, $\pi$).

The details of the warped hole cylinders shown in Fig. 3a are sensitive to the precise height of the As above the Fe square planes in the unit cell. For the experimentally determined As position, there are three hole cylindrical sections. The outer most cylinder flares out significantly at $k_z = \pi$. The sizes of the various ellipsoids with the AF Brillouin zone are sensitive to the precise position of the As and the magnitude of the ordered moment. The hole and electron sections make equal contributions to the density of states (to within ~2%), as expected for a spin density wave, with a total of $N(E_F)$=0.75 eV$^{-1}$ per formula unit, yielding a Sommerfeld coefficient $\gamma$=1.9 mJmole$^{-1}$K$^{-2}$.

**References**

1. Lee, P. A. 2007 <http://arXiv.org/abs/0708.2115v2>.

2. Anderson, P. W. *et al.* 2004 *J. Phys.: Condens. Matter* **16**, R755.

3. Kampf, A. P. 1994 *Phys. Rep*. **249**, 219.

4. Kamihara, Y. *et al*. 2008 *J. Am. Chem. Soc*. **130**, 3296.

5. de la Cruz, C. *et al*. 2008 *Nature* **453**, 899.

6. Chen, G. F. *et al* 2008 <http://arXiv.org/abs/0803.3790>.

7. McGuire M. A. *et al*. 2008 <http://arXiv.org/abs/0804.0796>.

8. Chen, G. F. *et al*. 2008 <http://arXiv.org/abs/0806.2648>.

9. Chen, T. Y. *et al*. 2008 *Nature* **453**, 1224.

10. Hunte, F. *et al*. 2008 *Nature* **453**, 903.

11. Rotter, M., Tegel, M., Johrendt, D. 2008 <http://arXiv.org/abs/0805.4630v1>.

12. Takahashi, H. *et al.* 2008 Nature **453**, 376.

13. Krellner, C. *et al.* 2008 <http://arXiv.org/abs/0806.1043>.

14. Steglich, F., *et al*. 1979 *Phys. Rev. Lett*. **43**, 1892.

15. Maple, M. B. *et al*. 1986 *Phys. Rev. Lett*. **56**, 185.

16. Hu, W. Z. *et al*. 2008 <http://arXiv.org/abs/0806.2652>.

17. Yan, J.-Q. et al. 2008 <http://arXiv.org/abs/0806.2711>.

18. Sebastian, S. E. et al. 2008 <http://arXiv.org/abs/0807.1896>.

19. Chen, G. F. *et al*. 2008 <http://arXiv.org/abs/0806.1209>.

20. Yelland, E. A. *et al*. 2007 *Phys. Rev. Lett.* **100**, 047003.

21. Sebastian, S. E. *et al*. 2008 *Nature* **454**, 200.

22. Shoenberg, D. 1984 *Magnetic oscillations in metals* (Cambridge University Press, Cambridge).

23. Dong, J. *et al*. 2008 <http://arXiv.org/abs/0803.3426>.

24. Singh, D. J. & Du, M. H. 2008 *Phys. Rev. Lett.* **100**, 237003.

25. Liu, H. *et al.* 2008 <http://arXiv.org/abs/0806.4806v1>.

26. Zhao, J. *et al.* 2008 <http://arXiv.org/abs/0807.1077>.

27. Fawcett, E. 1988 *Rev. Mod. Phys.* **60**, 209-283.

28. Ronning, F. *et al.* 2008 <http://arXiv.org/abs/0806.4599>.

29. Choi, T.-P. & Phillips, P. 2005 *Phys. Rev. Lett.* **95**, 196405.

30. Canfield, P. C., Fisk, Z. 1992 *Phil. Mag. B* **65**, 1117.

31.  Craven, J. E. 1969 *Phys. Rev.* **182**, 693.

32. Liu, C. *et al.* 2008 <http://arXiv.org/abs/0806.3453>.